\documentclass[nohyperref]{article}

    \PassOptionsToPackage{numbers, compress}{natbib}

\usepackage[preprint]{neurips_2023}

\usepackage{natbib}
\usepackage{tabularx}
\usepackage{amsmath}
\usepackage{amssymb}
\usepackage{mathtools}
\usepackage{amsthm}
\usepackage{booktabs}
\usepackage{subcaption}
\usepackage{paralist}
\usepackage[inline]{enumitem}
\usepackage{listings}
\usepackage{paralist}

\usepackage{soul}
\usepackage{color}
\usepackage{wallpaper}
\usepackage{graphicx}
\usepackage{numprint}
\usepackage{paralist}
\usepackage{xspace}
\usepackage{xcolor}

\usepackage[some]{background}
\usepackage[11pt]{moresize}
\usepackage{t1enc}

\definecolor{orangeN}{rgb}{1,.5,0}
\definecolor{blueN}{rgb}{.2, .59, .88}
\definecolor{purpleN}{rgb}{.294118, 0, .509804}
\definecolor{greenN}{rgb}{.421, .578, .241}
\definecolor{pinkN}{cmyk}{0, 0.7808, 0.4429, 0.1412}
\definecolor{grayN}{gray}{0.6}

\newcommand{\mila}[1] {{\color{purpleN} \bf{[mila here]}}} 
 
\newcommand{\humepros}{\textsc{Hume-Prosody}}
\newcommand{\humevb}{\textsc{Hume-VocalBurst}}

\newcommand{\modulatesemo}{\textsc{Modulate-Sonata}}
\newcommand{\modulategame}{\textsc{Modulate-Stream}}

\usepackage[utf8]{inputenc} 
\usepackage[T1]{fontenc}    
\usepackage{hyperref}       
\usepackage{url}            
\usepackage{booktabs}       
\usepackage{amsfonts}       
\usepackage{nicefrac}       
\usepackage{microtype}      
\usepackage{xcolor}         

\title{\Large{The NeurIPS 2023 Machine Learning for Audio Workshop:}\\ 
\Large{\textit{Affective Audio Benchmarks and Novel Data}}}

\author{%
 Alice Baird* \\
  Hume AI\\
  New York, USA\\
  \texttt{alice@hume.ai} \\
  \And
 Rachel Manzelli*\\
  Modulate AI\\
  Massachusetts, USA\\
  \texttt{rachel@modulate.ai } \\
    \And
         Panagiotis Tzirakis\\
   Hume AI\\
  New York, USA\\
  \And
     Chris Gagne\\
   Hume AI\\
  New York, USA\\
    \AND
     Haoqi Li\\
   Hume AI\\
  New York, USA\\
      \And
     Sadie Allen\\
   Boston University\\
   Boston, USA\\
  \And
     Sander Dieleman\\
   DeepMind\\
  London, UK\\
  \AND
     Brian Kulis\\
   Boston University\\
   Boston, USA\\
  \And
     Shrikanth S. Narayanan\\
   USC-SAIL\\
  California, USA\\
    \And
     Alan Cowen\\
   Hume AI\\
  New York, USA\\
}

\begin{document}

\maketitle

\begin{abstract}
The NeurIPS 2023 Machine Learning for Audio Workshop brings together machine learning (ML) experts from various audio domains. There are several valuable audio-driven ML tasks, from speech emotion recognition to audio event detection, but the community is sparse compared to other ML areas, e.g., computer vision or natural language processing. A major limitation with audio is the available data; with audio being a time-dependent modality, high-quality data collection is time-consuming and costly, making it challenging for academic groups to apply their often state-of-the-art strategies to a larger, more generalizable dataset. In this short white paper, to encourage researchers with limited access to large-datasets, the organizers first outline several open-source datasets that are available to the community, and for the duration of the workshop are making several propriety datasets available. Namely, three vocal datasets, \humepros{}, \humevb{}, an acted emotional speech dataset \modulatesemo, and an in-game streamer dataset \modulategame. We outline the current baselines on these datasets but encourage researchers from across audio to utilize them outside of the initial baseline tasks.
\end{abstract}

\section{Introduction}

Working with audio data in machine learning presents unique challenges compared to fields like computer vision. Despite the importance of various key problems in the audio domain, such as text-to-speech, voice recognition, source separation, and synthesis, it has received considerably lower attention. However, there has been a recent renaissance in audio research, particularly in the field of synthesis, with the release of several influential papers in the last year \cite{audiolm, musiclm, moûsai, audioLDM, msanii, wang2023neural}. The relative scarcity of prior research and this recent boom serves as the primary motivations behind organizing the 2023 NeurIPS Machine Learning for Audio (MLA) Workshop.

The workshop covers a vast scope of audio-related tasks, including but not limited to speech modeling, speech generation, music generation, denoising of speech and music, data augmentation, acoustic event classification, transcription, source separation, and even multimodal modelling involving audio.

There are prestigious competitions like the Detection and Classification of Acoustic Scenes and Events (DCASE) focusing on audio-driven machine learning \cite{mesaros2017dcase}, and numerous large-scale audio datasets with high-level labeling available in the literature, including Audioset~\cite{gemmeke2017audio}, Urban80k~\cite{salamon2014dataset}, and Librispeech \cite{panayotov2015librispeech}, along with unique concepts such as bird identification~\cite{kahl2022overview}, and music genre detection~\cite{sturm2012analysis}.
Despite the availability of such datasets, there still exists a scarcity of openly accessible large-scale datasets particularly tailored for more specialized domains, such as human-computer interaction and human behavior analysis. 
The lack of specialized datasets presents a considerable hurdle for developing systems that understand more nuanced human characteristics.

To address this issue and to support authors submitting their work to MLA, the organizers have proactively provided four datasets to researchers participating in the workshop, namely, the \humepros{}, \humevb{}, \modulategame{} and \modulatesemo{}. These datasets offer a broad spectrum of human states, including both labeled and unlabeled data.
Researchers are given unrestricted access to use these datasets for their submissions, providing them with the opportunity to tailor their strategies to new data situations and investigate innovative solutions.
Moreover, in addition to utilizing the datasets for applied research, the workshop encourages researchers to test their approaches against established benchmarks on baseline tasks outlined in previous machine learning competitions \cite{baird2022icml, baird2022acii}.

This paper aims to provide a comprehensive description of the four large-scale datasets made available in conjunction with this workshop (see Section \ref{sec:propri}). Additionally, for the datasets where applicable, we present benchmark results achieved thus far (see Section \ref{sec:baselines}).

\section{Workshop Audio Datasets}
\label{sec:propri}

Researchers have harnessed an array of audio datasets to train models in various disciplines, including automatic speech recognition and speech diarization, among others. However, there exists a conspicuous scarcity of large-scale datasets specifically designed for speech emotion recognition and the study of people's responses to events. To bridge this gap, our workshop presents substantial datasets derived from individuals interacting with video games, making them ideally suited for investigating human responses to dynamic events. We further provide labeled datasets for emotion recognition, focusing on speech prosody and vocal bursts. These resources offer an invaluable benchmark for the evaluation and enhancement of unsupervised models, marking a significant advancement in the field of speech emotion recognition research. In what follows, we present four datasets: \humepros{} (Sec. \ref{hume_pros}), \humevb{} (Sec. \ref{hume_vb}), \modulatesemo{} (Sec. \ref{modulate_emo}), and \modulategame{} (Sec. \ref{modulate_game})






\subsection{The Hume Expressive Prosodic Speech Dataset (\humepros{})}
\label{hume_pros}
\humepros{} represents a subset of a comprehensive dataset containing emotionally rated spoken utterances with diverse prosody. This subset comprises 41 hours, 48 minutes, and 55 seconds of audio data collected from 1,004 speakers, aged between 20 to 66 years old. The data was collected across three countries with distinct cultures: the United States, South Africa, and Venezuela. Notably, the data was recorded "in-the-wild", meaning it was captured in uncontrolled recording conditions using the speakers' own microphones.

The foundation of this dataset consists of more than 5,000 "seed" samples, which encompass various emotional expressions. These seed samples were gathered from openly available datasets such as MELD \cite{poria2018meld} and VENEC \cite{Laukka10-PTV,elfenbein2022we,laukka2016expression}. The seeds include a mix of 'same' sentences, such as over 500 instances of the phrase "Let me tell you something" \cite{Laukka10-PTV}, where the prosody plays a significant role in conveying meaning, and 'different' sentences, each with varying words and semantics, where the prosody's functional load is relatively lower.

Each audio sample in \humepros{} is associated with intensity labels for ten different expressed emotions, ranging from 1 to 100. The complete Hume-Prosody dataset comprises 48 emotional expression dimensions, based on the semantic-space model for emotion \cite{cowen2021semantic}. However, for this particular subset, which was introduced in this year's Computational Paralinguistic Challenge (ComParE)~\cite{schuller2023acm}, nine emotional classes were selected due to their more balanced distribution across the valence-arousal space. These classes include 'Anger,' 'Boredom,' 'Calmness,' 'Concentration,' 'Determination,' 'Excitement,' 'Interest,' 'Sadness,' and 'Tiredness.'

To create \humepros{}, participants were recruited through various crowdsourcing platforms like Amazon Mechanical Turk, Clickworker, Prolific, Microworkers, and RapidWorker. They were instructed to mimic a seed vocal burst they heard and use their computer microphone to record themselves imitating the seed sentence with similar prosody to the original recording. Each participant completed 30 trials per survey, and they could complete multiple versions of the survey. The study received informed consent from all participants, and its design and procedure were approved by Heartland IRB.

The intensity ratings for each emotion were normalized to a scale ranging from 0 to 1. For baseline experiments, the audio files were normalized to -3 decibels and converted to 16 kHz, 16-bit, mono format (the raw unprocessed audio is also provided, captured at 48 kHz). No additional processing was applied to the files, making the data amenable to various tasks. Subsequently, the data was divided into training, validation, and test sets, ensuring speaker independence, see \ref{tab:hume-pros} for an overview.

\begin{table*}[h]
\caption{Summary of the \humepros{} dataset, first presented at ComParE 2023~\cite{schuller2023acm}. Including number of samples, speakers, and distribution of identified gender. Test split remains blind.}
\label{tab:db}
\centering
\begin{tabular}{lrrr}
\toprule
 & \textbf{Train} & \textbf{Dev.} & \textbf{Test} \\
\midrule
\textbf{Sample no.} & 30,133 & 12,241 & - \\
\textbf{Speaker no.} & 600 & 202 & - \\
\textbf{Gender (f:m)} & 379:221 & 117:85 & - \\
\bottomrule
\label{tab:hume-pros}
\end{tabular}
\end{table*}

\subsection{The Hume Expressive Vocal Bursts Dataset (\humevb{})}
\label{hume_vb}
The \humevb{} dataset is a vast collection of emotional non-linguistic vocalizations, also known as vocal bursts. It includes audio data totaling 36 hours, 47 minutes, and 04 seconds (HH:MM:SS) from 1,702 speakers aged between 20 and 39 years. The dataset was compiled in four countries with diverse cultures: China, South Africa, the U.S., and Venezuela. In these recordings, individuals imitated expressive seed samples, showcasing various emotions. The speakers' vocalizations were recorded in the comfort of their homes using their microphones.

Each vocal burst in the dataset has been rated by an average of 85.2 raters for the intensity of 10 different expressed emotions. These emotions are Amusement, Awe, Awkwardness, Distress, Excitement, Fear, Horror, Sadness, Surprise, and Triumph, each rated on a scale from 1 to 100. The intensity ratings for each emotion were scaled to a range of 0 to 1.

For the baseline experiments, the audio files were standardized by normalizing them to -3 decibels and then converted to a 16 kHz, 16-bit, mono format. Additionally, the dataset also provides participants with the original, unprocessed audio, which was captured at 48 kHz. To ensure fair evaluation, the data was divided into training, validation, and test sets, taking into account speaker independence and maintaining a balance across different emotion classes. (Refer to Table \ref{tab:hume-vb} for the details of the data partitioning.)

\begin{table}[h]
\centering
\caption{Summary of the \humevb{} data, first presented at ExVo 2022~\cite{baird2022icml}. Including number of samples, speakers, and distribution of identified gender. The age range for speakers is 20.5-39.5 years. For the purposes of the competition, the test split remains blind, speakers given for few-shot task.}
\begin{tabular}{lrrr}
\toprule
             & \textbf{Train}    & \textbf{Dev.}      & \textbf{Test}  \\
             \midrule
\textbf{Sample no.}      & 19\,990   & 19\,396  & 19\,815   \\
\textbf{Speaker no.}     &   571    &   568    &   563   \\
\textbf{Gender (f:m)}          & 305:266  & 324:244  &   ---    \\
\bottomrule
\end{tabular}
\label{tab:hume-vb}
\end{table}

\subsection{The Modulate Acted Expression Speech Dataset (\modulatesemo)}
\label{modulate_emo}
\modulatesemo{} is an acted speech dataset, provided by Modulate. The dataset consists of 23 unique professional voice actors performing 15 total roles  (more than 6 hours of audio data). Each \verb|mp3| file contains the full audio recording of the actor speaking from a script of emotional sentences, in a particular \textit{style} (role). The audio files contain the full script alongside a text file of the emotional sentence spoken at that specific timestamp. 

There are 25 unique emotion classes in the dataset, including:

\begin{inparaitem}[,] `\textit{adoration}' \item `\textit{amusement}' \item `\textit{anger}' \item `\textit{awe}' \item `\textit{confusion}' \item `\textit{contempt}' \item `\textit{contentment}' \item `\textit{desire}' \item `\textit{disappointment}' \item `\textit{disgust}' \item `\textit{distress}' \item `\textit{elation}' \item `\textit{embarrassment}' \item `\textit{fear}' \item `\textit{hype}' \item `\textit{interest}' \item `\textit{pain}' \item `\textit{realization}' \item `\textit{relief}' \item `\textit{sadness}' \item `\textit{seductionecstasy}' \item `\textit{surprisenegative}' \item `\textit{surprisepositive}' \item `\textit{sympathy}' \item `\textit{triumph}'\end{inparaitem}. 

The 15 roles performed by the actors include anime voice archetypes (AVA), well-known actors, and fantasy characters. Specifically, these are: 

\begin{inparaitem}[,]
`\textit{AVABubblyAndSweet}' \item `\textit{AVANasallyAndMidpitched}'
\item `\textit{AVAChild}' \item `\textit{BatmanImpression}'
\item `\textit{EmmaWatsonImpression}' \item `\textit{JudyDenchImpression}'
\item `\textit{KeanuImpression}' \item `\textit{KristenBellImpression}'
\item `\textit{MatureAndSmokey}' \item `\textit{Nobility}'
\item `\textit{Outlander}' \item `\textit{Pirate}'
\item `\textit{ScarlettJohansonImpression}' \item `\textit{SmallCompanion}'
\item `\textit{Spellcaster}'
\end{inparaitem}.

The script of emotional sentences read by the actors is provided in Appendix~\ref{script}. For each emotional class there are between 2 to 6 sentences. 

For the purposes of benchmarking, we segment the long-form audio files following the emotion labels. This results in 756 total audio samples, with an average of 29.52 seconds in length, and between 29-32 segments for each class. Proceeding this we have prepared a speaker-independent partition split (see Table~\ref{tab:modulate-emo} for further details), based on the individual segments of audio. 

The audio was recorded using predominantly Neummann U87, SM7B, and Blue Yeti microphones, with some additional unknown condenser microphones used by actors who recorded their sessions remotely.

\begin{table}
\centering
\caption{Summary of the \modulatesemo{} data.  Including number of samples, speakers, and duration of audio data per partition split.}
\begin{tabular}{l r r r}
\toprule
 & \textbf{Train} & \textbf{Dev.} & \textbf{Test} \\
\midrule
\textbf{Sample no.} &  516 & 125  & 115  \\ 
\textbf{Speaker no.} & 16 & 4 & 3 \\
\textbf{Duration (HH:MM:SS)} & 04:11:03 & 00:59:04 & 01:01:54 \\
\bottomrule
\end{tabular}
\label{tab:modulate-emo}

\end{table}

\subsection{The Modulate Streamer Dataset (\modulategame)}
\label{modulate_game}
\modulategame{} is an audio-only dataset of over 7\,000 hours of publicly available gaming streams. The dataset is in total is 379GB in size, and contains 1\,880\,260 audio files saved in \verb|opus| format for space efficiency.

Metadata is contained in a separate TSV file, and contains the following attributes for each audio file: \begin{lstlisting}
[clip_name, player_id, session_id, clip_duration_msec, transcript,
num_words, game_metadata]
\end{lstlisting}
Each attribute is described as follows:
\begin{itemize}
  \item \verb|clip_name|: the audio filepath.
  \item \verb|player_id|: an anonymized ID of the speaker. There are 940 unique players in this dataset.
  \item \verb|session_id|: an anonymized ID of the stream. There are 2\,839 unique streams in this dataset.
  \item \verb|clip_duration_msec|: the clip duration in milliseconds.
  \item \verb|transcript|: the approximate transcription of the audio contained within the clip, estimated by a \verb|wav2vec|-based STT model.
  \item \verb|num_words|: the number of words in the approximated transcription, with an average value of 20 for this dataset.
  \item \verb|game_metadata|: the game the speaker in the clip is playing in their stream (or if not a game, the general topic of the stream), if provided. There are 110 unique topics/games in this dataset.
\end{itemize}

\begin{table}
\centering
\caption{Summary of the \modulategame{} data.  Including number of samples and speakers per partition split.}
\begin{tabular}{l r r r}
\toprule
 & \textbf{Train} & \textbf{Dev.} & \textbf{Test} \\
\midrule
\textbf{Sample no.} &  1\,835\,242 & 38\,079  & 6\,939  \\ 
\textbf{Speaker no.} & 658 & 141 & 141 \\
\bottomrule
\end{tabular}
\label{tab:modulate-game}

\end{table}

\section{Current Baselines, and Machine Learning Tasks}
\label{sec:baselines}

For the \humepros{} and \humevb{} datasets, various benchmarks are available, encompassing a variety of machine learning tasks. For \modulatesemo{} we provide a simple baseline for speech emotion recognition to validate the efficacy of the datasets target domain. Due to the size of \modulategame{} we do not provide any baseline results, however we have provided speaker-independent partition splits which anyone using the data are welcome to utilize.

In this section, we will provide an overview of the original baselines established by the workshop organizers for each dataset. Additionally, whenever applicable, we will showcase the best results achieved through contributions to previous competitions held where this data was provided. As \humepros{} and \humevb{} were provided as part of previously held competitions, the test sets for each of these will remain blind, and those wishing to evaluate for the tasks described herein should send predictions to \url{competitions@hume.ai}. 

\subsection{\humepros{} Description of Tasks}

The \humepros{} have been assigned only one task, which is publicly available with detailed information provided by the ComParE challenge organizers~\cite{schuller2023acm}. You can find the baseline code for this task at \url{https://github.com/EIHW/ComParE2023}.

This specific task for the \humepros{} is known as the \textit{Emotion Share Sub-Challenge}, which involves a multi-label regression task. It requires participants to predict the proportion or 'share' of nine different emotions based on the ratings given by multiple raters for the 'seed' sample.

\subsection{\humepros{} Initial Baseline Results}

For this baseline, embeddings were extracted from Wav2Vec2, fine-tuned to the MSP-Podcast dataset\cite{Lotfian_2019_3}, to establish an initial baseline for the dataset. Next, a Support Vector Regressor was trained and evaluated, with the cost parameter $C$ of the SVM optimized based on performance on the Dev set. After this optimization process, a final model was trained using the concatenated training and Dev sets for evaluation on the Test partition.

\begin{table}[h!]
\caption{Results for \humepros{}. The official best results for Test 
 are emphasized; Reporting Pearsons Correlation Coefficient across the mean of the available classes ($\rho$).}
\centering
\begin{tabular}{lccc}
\toprule
($\rho$) & Dev. & Test & CI on Test \\
\midrule
Wav2Vec2 & .500 &\textbf{.514} &.499 – .529 \\
ComParE & .359 & .365 & .347 – .382 \\
Late Fusion & .470 & .476 & .461 – .492\\
\bottomrule
\end{tabular}
\label{tab:results-hpc}
\end{table}

The competition is at this time still on-going and so we would advise the authors to check the literature relating to the ComParE 2022 challenge for any updated benchmarks. 

\subsection{\humevb{} Description of Tasks}

The \humevb{} was first introduced during the ICML Expressive Vocal Burst (ExVo) Challenge~\cite{baird2022icml}, highlighting three key tasks with an emphasis on audio machine learning. Subsequently, it was presented at the Affective Vocal Burst (A-VB) Workshop at the 2022 ACII conference~\cite{baird2022acii}, where four additional tasks were defined with a focus on affective computing. In this section we will briefly describe each but please see the reference papers for further information. 

\textbf{ExVo Multi-Task Learning:}
Participants in this track will train multi-task models to predict 10 emotions, the speaker's age, and native country using vocal bursts. The baseline performance metric, is a combined metric (see equation ~\ref{eq:mtl}) based on mean Concordance Correlation Coefficient (CCC) for emotions, Mean Absolute Error (MAE) for age, and Unweighted Average Recall (UAR) for native country, all of which will determine the final standings.

    \begin{equation}
        \mathrm{S_{MTL}} = \frac{3}{(1/\mathcal{\hat{C}} + 1/\mathcal{\hat{M}} + 1/\mathcal{\hat{U}})}.
        \label{eq:mtl}
    \end{equation}
    
\textbf{ExVo Emotion Generation:}
This track requires teams to train generative models to produce vocal bursts for 10 distinct emotions. For evaluation in the competition combine metric was presented which included the quantitative methods of Fréchet Inception Distance (FID)~\cite{heusel2017gans}: 

    \begin{equation}
        \mathrm{FID} =\|\mu - \mu^*\|^2 + Tr( C + C^* - 2(C  C^*)^{1/2}), 
        \label{eq:fid}
    \end{equation}

 As well as a qualitative approach based on human ratings (HEEP): 
 
 \begin{equation}
    \mathrm{HEEP} = \sigma_{TH}/\sqrt{\sigma_T^2\sigma_H^2 },
    \label{eq:heep}
\end{equation}
    where $T$ corresponds to the vectorized target matrix, a dummy matrix of size $N$ (number of generated vocal bursts) by 10 (emotions), with ones for targeted emotions and zeros for non-targeted emotions, and $H$ corresponds to the vectorized rating matrix, a matrix of size $N\times10$ with entries corresponding to the average human intensity ratings of each generated vocal burst.

The two are then combined as $S_{GEN}$, and compute the mean between the inverted FID distance, and the HEEP score for each emotion ($e$); this is defined as 

        \begin{equation}
            S_{GEN}{_e}={\frac {1/FID_e+HEEP_e}{2}}.
            \label{eq:gen}
        \end{equation}

Please note the human evaluation was provided by the organizer for the ExVo competition only. 
    
\textbf{ExVo Few-Shot Emotion Recognition:}
In this innovative track, teams will recognize emotional vocalizations using few-shot learning, emphasizing personalization by considering factors like pitch and frequency of the speaker's voice. Two labeled samples per speaker will be used for personalization.

\textbf{A-VB High-Dimensional Emotion:}
Participants will predict the intensity of 10 specific emotions through a multi-output regression approach, evaluating their performance using the mean CCC across all emotions.

\textbf{A-VB Two-Dimensional Emotion:}
Focused on predicting arousal and valence from the circumplex model of affect~\cite{russell1980circumplex}, participants will approach this task as a regression problem, reporting performance using the mean CCC across both dimensions.

\textbf{A-VB Cross-Cultural Emotion:}
This unique track introduces a 10-dimensional emotion intensity regression task for four different countries, challenging participants to predict the intensity of 40 emotions. The evaluation will use the mean CCC across all 40 emotions.

\textbf{A-VB Expressive Burst-Type:}
For this classification task, participants will aim to classify eight types of expressive vocal bursts (e.g., Laugh, Cry, Scream). Performance will be assessed using the Unweighted Average Recall (UAR), serving as an accuracy measure.

\subsection{\humevb{} Initial Baseline Results}

For \humevb{} there are several baselines set for the various  tasks described. For both competitions the baseline code is provided at \url{https://github.com/HumeAI/competitions}. 

For each competitions a variety of approaches were presented, including feature-driven~\cite{eyben2010opensmile}, end-to-end~\cite{tzirakis2017end} Long-short-term memory-based architectures, and Generative Adversarial Networks for generation. 

\begin{table*}[h]
\centering
\caption{Results for the 2022 ICML ExVo Workshop~\cite{baird2022icml}. Reporting best results for ExVo MultiTask, ExVo Generation, and ExVo FewShot Tasks.}
\label{tab:combined_results}

\subcaption{Development and baseline test scores for ExVo-MultiTask, reporting best for each as given in~\cite{baird2022icml}.}
\begin{tabular}{l | r r r c | c}
\toprule
            & \multicolumn{4}{c |}{\textbf{Development}}       & \textbf{Test}     \\
            & E-CCC  & C-UAR  & A-MAE  & $S_{MTL}$      & $S_{MTL}$         \\
\midrule
ComParE        & \textbf{.416}   & \textbf{.506}   & 4.22     & \textbf{.349} $\pm$ .003  & \textbf{.335} $\pm$ .002 \\
eGeMAPS        &  .353           & .423            & \textbf{4.01}   & .324 $\pm$ .005           & .314 $\pm$ .005 \\
\bottomrule
\end{tabular}

\vspace{0.5cm}

\subcaption{Fr\'{e}chet inception distance (FID) for the ExVo-Generation task, for each emotions, no baseline was given for `Triumph' (Tri.), as no samples were generated for this class.}
\begin{tabular}{l | r r r r r r r r r r}
\toprule
 & Amu. & Awe & Awk. & Dis. & Exc. & Fea. & Hor. & Sad. & Sur. & Tri. \\
\midrule
FID & 4.92 & 4.81 & 8.27 & 6.11 & 6.00 & 5.71 & 5.64 & 5.00 & 6.08 & -- \\
HEEP & 0.49 & 0.46 & 0.04 & 0.32 & 0.08 & 0.04 & 0.27 & -0.03 & 0.22 & -- \\
$S_{GEN}$ & 0.345 & 0.33 & 0.08 & 0.24 & 0.13 & 0.11 & 0.22 & 0.08 & 0.19& 0.00 \\
\bottomrule
\end{tabular}

\vspace{0.5cm}

\subcaption{Test results for the ExVo FewShot task. Reporting best score given in~\cite{baird2022icml}.}
\begin{tabular}{l | r r r r r r r r r r | r}
\toprule
$\mathcal{H}$
  & Amu. & Awe & Awk. & Dis. & Exc. & Fea. & Hor. & Sad. & Sur. & Tri. & $\mathcal{\hat{C}}$ \\ 
  \midrule
256 & .554 & .581 & .282 & .420 & .311 & .544 & .490 & .383 & .561 & .315 & $\textbf{.444} \pm .006$ \\
\bottomrule
\end{tabular}

\end{table*}

\begin{table*}
    \centering
    \caption{Baseline scores for A-VB 2022~\cite{baird2022icml}. Reporting the mean Concordance Correlation Coefficient (CCC) for the three regression tasks and the Unweighted Average Recall (UAR) across the 8-classes for the vocal burst type task.}
    \begin{tabular}{l | rr rr rr | rr}
        \toprule
        & \multicolumn{6}{c|}{CCC} & \multicolumn{2}{c}{UAR} \\
        & \multicolumn{2}{c}{\textit{High}} & \multicolumn{2}{c}{\textit{Two}} & \multicolumn{2}{c|}{\textit{Culture}} & \multicolumn{2}{c}{\textit{Type}} \\
        & Dev. & Test & Dev. & Test & Dev. & Test & Dev. & Test \\
        \midrule
        \textit{ComParE}~\cite{eyben2010opensmile} & .515 & .521 & .49¢ & .499 & .387 & .380 & .391 & .384 \\
        \midrule
        \textsc{End2You}~\cite{tzirakis2017end} & .564 & \textbf{.569} & .499 & \textbf{.508} & .436 & \textbf{.440} & .417 & \textbf{.417} \\
        \bottomrule
    \end{tabular}

    \label{tab:results}
\end{table*}

\subsection{\humevb{} Latest Benchmark Results}

As the competitions related to \humevb{} we can provide the full list of results here for each, where applicable citing the relevant literature for the approach. 
\begin{table*}[htbp]
\centering
\caption{Latest benchmarks for \humevb{} based on the tasks presented in the 2022 ExVo Workshop.  Best known result on the test set is emphasized. Reporting, $S_{MTL}$ (see equation \ref{eq:mtl}), $S_{Gen}$ (see equation \ref{eq:gen}) and Concordance Correlation Coefficient (CCC), the task type respectively.}
\label{tab:exvo_results}
\begin{tabular}{l l l r}
\toprule
\textbf{Team}             & \textbf{Task}          & \textbf{Metric} & \textbf{Test} \\
\midrule
Organisers~\cite{baird2022icml}       & MultiTask         & $S_{MTL}$ & .335 \\
TeamAtmaja~\cite{atmaja2022jointly}                & --                    & -- & .378 \\
IdiapTeam~\cite{purohit2022comparing}                & --                    & -- & .379 \\
EIHW-MM~\cite{jing2022redundancy}                 & --                    & -- & .394 \\
0xAC~\cite{belanich2022multitask}                      & --                    & -- & .407 \\
CMU\_MLSP~\cite{sharma2022selfsupervision}                 & --                    & -- & .412 \\
NLPros~\cite{anuchitanukul2022burst2vec}                    & --                    & -- & \textbf{.435} \\

\midrule
Organisers~\cite{baird2022icml}       & Generate          & $S_{GEN}$ & .090 \\
Resemble~\cite{hsu2022synthesizing}                 & --                    & -- & .119 \\
StyleMelMila~\cite{jiralerspong2022generating}            & --                    & -- & \textbf{.408} \\
\midrule
Organisers~\cite{baird2022icml}       & FewShot           & CCC & .444 \\
EIHW-MM~\cite{jing2022redundancy}                 & --                    & -- & .650 \\
SaruLab-UTokyo~\cite{xin2022exploring}           & --                    & --&\textbf{ .739} \\

\bottomrule
\end{tabular}
\end{table*}

\begin{table}[htbp]
\centering

\caption{Latest benchmarks for \humevb{} based on the tasks presented in the 2022 A-VB Workshop. Best known result on the test set is emphasized. Report Concordance Correlation Coefficient (CCC), and Unweighted Average Recall (UAR) for the regression and classification tasks, respectively.}
\begin{tabular}{l l l l}
\toprule
\textbf{Team} & \textbf{Task} & \textbf{Metric} & \textbf{Test} \\
\hline
\midrule
Organizers~\cite{baird2022acii} & High & CCC & .569 \\
TeamEP-ITS~\cite{atmaja2022predicting} & -- & -- & .655 \\
SclabCNU~\cite{nguyen2022finetuning} & -- & -- & .668 \\
HCAI \cite{hallmen2022efficient} & -- & -- & .685 \\
HCCL~\cite{li2023hierarchical} & -- & -- & .724 \\
Anonymous~\cite{trinh2022selfrelation} & -- & -- & .730 \\
EIHW~\cite{karas2022selfsupervised}  & -- & -- & \textbf{.736} \\
\hline
Organizers~\cite{baird2022acii} & Two & CCC & .508 \\
SclabCNU~\cite{nguyen2022finetuning}  & -- & -- & .620 \\
TeamEP-ITS~\cite{atmaja2022predicting} & -- & -- & .629 \\
HCCL~\cite{li2023hierarchical} & -- & -- & .685 \\
EIHW~\cite{karas2022selfsupervised}  & -- & -- & \textbf{.707} \\
\hline
Organizers~\cite{baird2022acii} & Culture & CCC & .440 \\
TeamEP-ITS~\cite{atmaja2022predicting} & -- & -- & .520 \\
HCAI \cite{hallmen2022efficient} & -- & -- & .526 \\
SclabCNU~\cite{nguyen2022finetuning}  & -- & -- & .550 \\
HCCL~\cite{li2023hierarchical} & -- & -- & .602 \\
EIHW~\cite{karas2022selfsupervised}  & -- & -- &\textbf{.620} \\
\hline
Organizers~\cite{baird2022acii} & Type & UAR & .417 \\
TeamEP-ITS~\cite{atmaja2022predicting} & -- & -- & .490 \\
SclabCNU~\cite{nguyen2022finetuning}  & -- & -- & .497 \\
EIHW~\cite{karas2022selfsupervised} & -- & -- & .562 \\
Team-AVB~\cite{syed2022classification} & -- & -- & .519 \\
HCAI~\cite{hallmen2022efficient} & -- & -- & \textbf{.586} \\
\bottomrule
\bottomrule
\end{tabular}
\end{table}

\subsection{\modulatesemo{} Description of Tasks}
Given the high quality of the audio, and the scenario being acted, we suggest that the \modulatesemo{} data be used for generation or as a benchmark for evaluation of speech emotion recognition tasks. With this in mind, we provide an initial speech emotion classification task utilizing all classes. 

\subsection{\modulatesemo{} Initial Baseline Results}

For \modulatesemo{} we provide an initial baseline for the partitions provided. We extracted HuBERT and Wav2Vec2 embeddings utilizing the default parameters from each of the audio files, and took the mean across the time axis. For a classifier we utilize the \verb|sklearn| Logistic Regression module, and optimize the value for C=\{0.001, 0.01, 0.1, 1\} on the validation set, and retrain with the model with the validation and training set concatenated using on the best value for C, evaluated on the test set. 

From the results in Table \ref{tab:model_performance} we can see that Wav2Vec2 and HuBERT embeddings perform similarly for the emotion recognition task, and fusing these embeddings shows further improvement. In all cases,  the strong performance across the data indicates the validity of this data for both speech emotion recognition and generation type tasks.  

\begin{table}[h!]
\centering
\caption{\modulatesemo{} validation and test sets performance, for 25 class, speech emotion recognition reporting Unweighted Average Recall (UAR), chance level 0.04.}
\begin{tabular}{l r r }
\toprule
\textbf{Embedding} & \textbf{Validation} & \textbf{Test} \\
\midrule
Wav2Vec2 & 0.80 & 0.85 \\
HuBERT & 0.79 & 0.86 \\
\midrule
early fusion & 0.82 & 0.90 \\
\bottomrule
\end{tabular}

\label{tab:model_performance}
\end{table}

\section{Summary and Conclusions} 
\label{sec:conclusions} 

In conclusion, for the 2023 NeurIPS Audio for Machine Learning Workshop we have made efforts to addresses the scarcity of specialized audio datasets by providing valuable resources, namely, \humepros{}, \humevb{}, \modulatesemo{}, and \modulategame{}. These datasets offer opportunities for researchers to explore diverse audio tasks, such as emotion recognition, vocal burst classification, speech generation, and unsupervised audio driven tasks. By establishing the outlined baselines and encouraging collaboration, the workshop aims fosters innovation in audio-driven machine learning, leading to the development of more robust models, advancing areas including human behavior understanding and human-computer-interaction amongst other.




\bibliography{main}
\bibliographystyle{unsrt} 

\section{Appendix}
\subsection{\modulatesemo{} Script}\label{script}

\subsubsection*{Amusement}
“I’m the Juggernaut, bitch.” “Want to know how I got these scars?...He turns to me and he says: Why. So. Serious. Let’s put a smile on that face!” “In the face of overwhelming odds, I’m left with only one option. I’m gonna have to science the shit outta this.”

\subsubsection*{Embarrassment}
“I didn’t mean to...just...just look away please!” “Are you kidding? I sounded like a total buffoon! I’m sure she’ll never even want to talk to me again.”

\subsubsection*{Sadness}
“They cursed us. Murderer they called us. The cursed us, and drove us away. And we wept, Precious, we wept to be so alone. And we only wish to catch fish so juicy sweet. And we forgot the taste of bread...the sound of trees...the softness of the wind. We even forgot our own name. My precious.” “Sometimes I wish I had never met you. Because then I could go to sleep at night not knowing there was someone like you out there.” “We’re going to be okay. You can rest now.”

\subsubsection*{Elation}
“I love the smell of napalm in the morning.” “Carpe diem. Seize the day, boys. Make your lives extraordinary!” “Whew! That’s the stuff!”

\subsubsection*{Triumph}
“Are you not entertained! Are you not entertained! Is this not why you are here?!” “King Kong ain’t got shit on me!” “We will not go quietly into the night! We will not vanish without a fight! We’re going to live on! We’re going to survive! Today we celebrate...our Independence Day!”

\subsubsection*{Contempt}
“My name is Maximus Decimus Meridius, commander of the Armies of the North, General of the Felix Legions and loyal servant to the true emperor, Marcus Aurelius. Father to a murdered son. Husband to a murdered wife. And I will have my vengeance, in this life or the next.” “You can torture us and bomb us and burn our districts to the ground. But do you see that? Fire is catching. And if we burn...you burn with us.” “I’m surrounded by idiots.”

\subsubsection*{Disgust}
“Get your stinking paws off me, you damned dirty ape.” “I can hardly forbear hurling things at him.”

\subsubsection*{Disappointment}
“Why is the rum gone? Why is the rum...always gone?” “Well look at you now; you just got your asses whipped, by a bunch of goddamn nerds. I’m sure your father is rolling his his grave.” “I’m not angry with you. I’m just...disappointed.”

\subsubsection*{Anger}
“You talkin’ to me? You talking to ME? ARE YOU TALKING TO ME? THEY CALL ME MR. PIG!” “Badges? We ain’t got no badges! We don’t need no badges! I don’t have to show you any
stinking badges!” “Enough is enough! I have had it with these motherfucking snakes on this motherfucking
plane!”

\subsubsection*{Adoration}
“When I drift off, I will dream about you. It’s always you.”
“Because I still wake up every morning...and the first thing I want to do is see your face.”

\subsubsection*{Sympathy}
“Ye best start believin’ in ghost stories, Miss Turner. Yer in one.” “Oh yes, the past can hurt. But you can either run from it, or learn from it.” “If I told you what happens...it won’t happen.”

\subsubsection*{Distress}
“You don’t understand! I coulda had class. I coulda been a contender. I could’ve been somebody, instead of a bum, which is what I am.” “I don’t know where I am. I don’t know what’s going no. I think I lost somebody but I..I can’t
remember...” “Gentlemen, you can’t fight in here! This is the War Room!” “Westballz doesn’t even try to win. He just styles on you as hard as possible, and he’s so good
at that, he just wins.”

\subsubsection*{Pain}
“I would have followed you...my brother...my captain...my king.” “I don’t care!” Harry yelled, snatching up a lunascope and throwing it into the fireplace. “I’ve had enough, I’ve seen enough, I want it to end, I don’t care anymore!” “But anyways, I just - ouch! Oh, sonnoffabitch, right on the corner of the table, ooh, that one stings...”

\subsubsection*{Hype}
“This is where we fight! This is where they die!” “Leeeeroy Jenkins!”

\subsubsection*{Seduction / Ecstasy}
“Mm...this chocolate is rich...and velvety...and so very delicious...” “I bet you would enjoy that luscious dragonfruit very much, wouldn’t you..” “Yes.. I want the cupcakes... I want all of the cupcakes.” “Your appetite... is very impressive....” “The yogurt! It flows like water. It’s soo good. I’ll have what I’m having!”

\subsubsection*{Desire}
“It is mine, I tell you. My own. My precious....yes, my precious...” “If I could just see her again...hear her again...touch her again...maybe I could die happy.”

\subsubsection*{Contentment}
“Mama always said life was like a box of chocolates. You never know what you’re gonna get.” “Elementary, my dear Watson.” “Get busy living, or get busy dying.” “No thanks, I’m perfectly happy sitting here with the sun on my face reading a good book for as long as the weather will let me.”

\subsubsection*{Relief}
“And...and I look at you, and I - I’m home.” “Gentlemen, I wash my hands of this weirdness.” “We’re...we’re alive? Oh, thank the merciful gods, we’re alive.”

\subsubsection*{Realization}
“Houston, we have a problem.” “You keep using that word. I do not think it means what you think it means.” “Oh. Oh balls.” “Oh. Em. Gee.” “Oh my god. Did...did that just happen?” “Ahah! An act of true love can heal a frozen heart!”

\subsubsection*{Interest}
“Hope. It is the only thing stronger than fear. A little hope is effective. A lot of hope is dangerous. A spark is fine, as long as it’s contained.” “What do you mean, you have a plan? Care to share?”

\subsubsection*{Confusion}
“The plot thickens, as they say. Why, by the way? Is it a soup metaphor?” “What is this? A center for ants?”

\subsubsection*{Awe}
“Toto, I’ve got a feeling we’re not in Kansas anymore.” “You’re gonna need a bigger boat.” “Yoo! Did he just walk up...slowly...and down smash?!”

\subsubsection*{Surprise (positive)}
“Cinderella story. Outta nowhere. A former greenskeeper, now, about to become the Masters champion. It looks like a mirac - It’s in the hole! It’s in the hole! It’s in the hole!” “And the crowd’s cheering for a four stock...this could be it...THIS IS IT!”

\subsubsection*{Surprise (negative)}
“You drilled a hole in the dentist?!” “What the - what the hell has been going on here? What, did you think I wouldn’t notice?”

\subsubsection*{Fear}
“I’m not going near there! Sunnyside is a place of ruin and despair, ruled by an evil bear who
smells of strawberries!” “You want me to fight? No way, man, he’s gonna kill me. He’s gonna kill me and my whole
family. I’m out, you hear me, I’m out!”

\end{document}